\documentclass{aa}
\usepackage{times}
\usepackage{epsfig}
\begin{document}
\thesaurus{01(03.13.4,12.07.1,08.16.2,08.02.3,12.04.1)}
\title{A robust and efficient method for calculating the magnification of extended sources
caused by gravitational lenses}
\author{M. Dominik\inst{1}\inst{2}\thanks{Work carried out at the Space Telescope
Science Institute is financed by a research grant from Deutsche
Forschungsgemeinschaft}}
\institute{Institut f\"ur Physik, Universit\"at Dortmund, D-44221 Dortmund, Germany \and Space Telescope Science Institute, 3700 San Martin Drive, Baltimore,
MD 21218, USA}
\date{Received ; accepted}
\titlerunning{Calculating the gravitational magnification of extended sources}
\maketitle
\begin{abstract}
To determine the magnification of an extended source caused by gravitational lensing
one has to perform a two-dimensional
integral over point-source magnifications in general. Since the point-source magnification
jumps to an infinite value on caustics, special care is required.
For a uniformly bright source, it has been shown earlier that the calculation 
simplifies if one determines the magnification from the area of the images of the extended
source by applying Green's theorem so that one ends up with a one-dimensional integration
over the image boundaries. This approach is discussed here in detail, and it is shown that
it can be used to yield a robust and efficient method also for limb-darkened sources.
It is also shown that the centroid shift can be calculated in a similar way.

\end{abstract}

\section{Introduction}
For fitting light curves for the ongoing microlensing events, there
is a need for robust and efficient methods for calculating the magnification of extended sources, which
are not limited to point-lenses.
Among the observed events, the presence of binary lenses is a 
reality (Dominik \& Hirshfeld~\cite{MLMC1let},\cite{MLMC1}; Udalski et al.~\cite{OGLE7}; 
Alard et al.~\cite{DUO2};
Bennett et al.~\cite{MLMC9}), and planetary events
involve a special case of a binary lens. 
In addition, for some configurations, the light curve for a limb-darkened source will differ significantly from that
of a uniformly bright source.  The limb-darkening effect has
recently been observed in the galactic microlensing event MACHO 97-BLG-28, which involves both an
extended source and a binary
lens, by the
PLANET collaboration (Albrow et al.~1998a,b\nocite{M28Planet1}\nocite{M28Planet2}); the fitting has been done by myself 
using the algorithm described in this letter. 

If one wants to integrate the point-source magnification 
in two dimensions one has to take special care of the position of the caustics, where the point-source magnification becomes 
infinite. While this integration can be performed easily for a point-mass lens (e.g. 
Schneider et al.~\cite{SEF}, p. 313; Witt \& Mao~\cite{WM}; 
Sahu~\cite{Sahu}; Dominik~\cite{DoDiss}),
this would be a difficult task for a general lens (e.g. a binary lens),
especially at a cusp singularity. In contrast, the area
of the images of the extended source and therefore its magnification remains continuous when the source hits a
caustic. The determination of the extended source magnification from the boundaries of the image areas has been
used to analyze the images of background galaxies behind
a cluster of galaxies 
(Dominik~\cite{DoDipl}).
The image boundaries can be obtained with a contour plot of an implicit function
describing the source boundary in the lens plane (Schramm \& Kayser~\cite{SK}). This method has been
expanded with routines for correcting, testing and finally analyzing the
contour line in order to produce an efficient and safe algorithm 
(Dominik~\cite{DoAstro}). In that paper, it is noted that it is easy to 
analyze the images from the contour line data, and an example is given, where quantities such as the area, 
width, length and curvature of the image have been determined.

Concerning microlensing light curves, it has been noted by Bennett \& Rhie (\cite{BenRhie}) that it is advantageous
to integrate in the lens plane rather than in the source plane to determine the magnification of an extended
source. For uniformly bright sources, Gould \& Gaucherel (\cite{GouGau}) 
proposed applying Green's theorem so that only one integration along the image boundary
must be performed rather than two over the
image area. This approach is identical to that used earlier (Dominik~\cite{DoDipl},~\cite{DoAstro}).
The contour plot method is the most convenient way to obtain data points on the image boundary from which the area
can be calculated.
In Sect.~2, this general approach is described,  
Sect.~3 gives details for a uniformly bright source, and Sect.~4 shows how this approach can also be used for
limb-darkened sources, in which case an easy-to-perform two-dimensional integration remains. 
In Sect.~5, the calculation of the centroid shift is discussed.

\section{Magnification and Green's theorem}
The magnification is given by the ratio of the area of the images to the area
of the source  
in the absence of the lens. Let $(y_1,y_2)$ denote arbitrary cartesian source coordinates and $(x_1,x_2)$ 
denote corresponding image 
coordinates, where in the absence of the lens $\vec y = \vec x$.
Let $I_\mathrm{s}(\vec y)$ denote the surface brightness of the source and
$I_\mathrm{l}(\vec x)$ denote the surface brightness of the images. The conservation of photon number then requires
(see e.g.~Schneider et al.~\cite{SEF}, p.~33; Schramm \& Kayser~\cite{SK}; Kayser \& Schramm~\cite{KS}) that
\begin{equation}
I_\mathrm{l}(\vec x) = I_\mathrm{s}(\vec y(\vec x))\,,
\label{brightlaw}
\end{equation}
where $\vec y(\vec x)$ gives the source position $\vec y$ related to the image position $\vec x$.
 
With $A_\mathrm{s}$ being the region in the $(y_1,y_2)$-plane subtended by the source and 
$A_\mathrm{l}$ being the region in the $(x_1,x_2)$-plane subtended by the images,
the magnification is given by
\begin{equation}
\mu = \frac{\int_{A_\mathrm{l}} I_\mathrm{l}(\vec x)\,\mathrm{d}x_1\,\mathrm{d}x_2}{\int_{A_\mathrm{s}} 
I_\mathrm{s}(\vec y)\,\mathrm{d}y_1\,
\mathrm{d}y_2}\,.
\label{magn0}
\end{equation}

In the limit of a point source, one obtains $\mu = \sum \tilde{\mu}(\vec x_j)$, where the sum runs over the
$m$ images $\vec x_j$ of the source, and $\tilde{\mu}$ is given by
\begin{equation}
\tilde{\mu}(\vec x_j) = \frac{1}{\left|\det\left(\frac{\partial \vec y}{\partial \vec x}\right)\left(\vec x_j\right)\right|}\,.
\label{magpt}
\end{equation}

The normalization of the brightness profile function $I_\mathrm{s}$ can be chosen so that the integral in the denominator (Eq.~(\ref{magn0})) becomes the
area of the source,
and for a circular source of radius $R_\mathrm{src}$ one obtains
\begin{equation}
\int_{A_\mathrm{s}} I_\mathrm{s}(\vec y)\,\mathrm{d}y_1\,\mathrm{d}y_2 = \pi R_\mathrm{src}^2\,.
\label{normalization}
\end{equation}

For a uniformly bright source, one has $I_\mathrm{s}(\vec y) = 1$ for positions within the source, so that
$I_\mathrm{l}(\vec x) = 1$ for positions within the images,
and the magnification becomes
\begin{equation}
\mu = \frac{1}{\pi R_\mathrm{src}^2}\,\int_{A_\mathrm{l}} \mathrm{d}x_1\,\mathrm{d}x_2\,,
\end{equation}
where the remaining integral is just the area of the images.

Green's theorem now states that for two functions $P(x_1,x_2)$ and $Q(x_1,x_2)$ which are
continous in a region $A$ and whose partial derivatives $\partial P/\partial x_2$ and
$\partial Q/\partial x_1$ are also continuous in $A$,
\begin{equation}
\int_{A} \left(\frac{\partial Q}{\partial x_1} - \frac{\partial{P}}{\partial x_2}\right)\,
\mathrm{d}x_1\,\mathrm{d}x_2 =
\int_{\partial A} P\,\mathrm{d}x_1 + Q\,\mathrm{d}x_2\,,
\end{equation}
where $\partial A$ denotes the boundary of the region $A$, which has to be piecewise-smooth.
Note that the values of $P$ and $Q$ outside the region $A$ do not play a role and can be chosen arbitrarily as
long as $P$ and $Q$ satisfy the conditions above.

\section{Uniformly bright sources}
\label{secub}
For a uniformly bright source, one can now choose $P = -\frac{1}{2} x_2$ and $Q = \frac{1}{2} x_1$ to obtain
\begin{equation}
\mu = \frac{1}{2 \pi R_\mathrm{src}^2}\,\int_{\partial A} x_1\,\mathrm{d}x_2 - x_2\,\mathrm{d}x_1\,.
\end{equation}

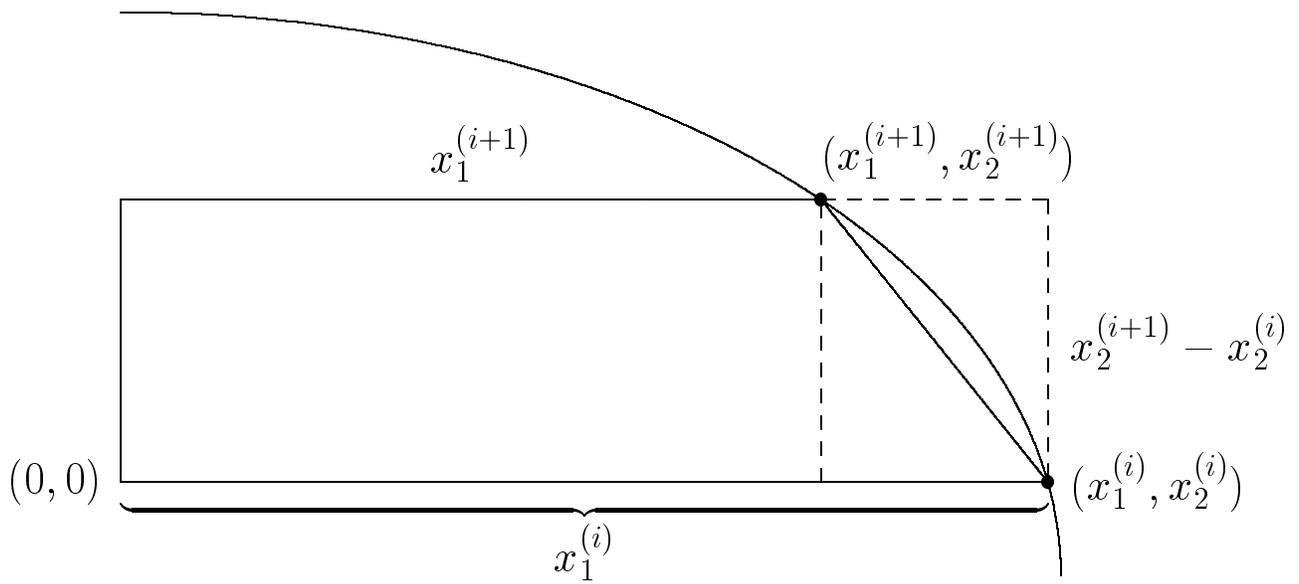
\begin{figure}
\vspace{4cm}
\caption{Two successive points $\vec x^{(i)}$ and $\vec x^{(i+1)}$ on the image boundary and the corresponding area segment}
\label{symfigure}
\end{figure}

For $n$ discrete positions $\vec x^{(i)}$ on the image boundary, where
$\vec x^{(n+i)} = \vec x^{(i)}$, which are obtained e.g.~by a contour plot, the magnification
can be approximated by
\begin{eqnarray}
\mu & = & \frac{1}{2 \pi R_\mathrm{src}^2}\,\sum_{i=1}^{n} \left[x_1^{(i)}\,(x_2^{(i+1)}-x_2^{(i)}) 
\,-\right.\nonumber \\ 
& & -\,\left.x_2^{(i)}\,(x_1^{(i+1)} - x_1^{(i)})\right]\,.
\end{eqnarray}
This means that one adds up rectangular segments with the length $x_1^{(i)}$ and the
width $\Delta x_2^{(i)} = x_2^{(i+1)} - x_2^{(i)}$ (and corresponding segments by interchanging the axes).
However $x_1^{(i+1)}$ has an equal ``right'' to be convolved with $\Delta x_2{(i)}$, yielding the  
symmetric version
\begin{eqnarray}
\mu & = & \frac{1}{4 \pi R_\mathrm{src}^2}\,\sum_{i=1}^{n} 
\left[(x_1^{(i)}+x_1^{(i+1)})\,(x_2^{(i+1)}-x_2^{(i)}) \,-\right. \nonumber \\
& & \left.-\,(x_2^{(i)}+x_2^{(i+1)})\,(x_1^{(i+1)} - x_1^{(i)})\right]\,,
\label{magnub}
\end{eqnarray}
which sums over trapezoidal segments having an area which is the mean of the rectangular areas formed from
segments
with length $x_1^{(i)}$ and $x_1^{(i+1)}$, and corresponds to a replacement of the true boundary
by a polygon with $n$ corners at $\vec x^{(i)}$ (see also Figure~\ref{symfigure}).
By increasing the number of points used, the magnification can be determined to the desired precision.

\section{Limb-darkened sources}
Let us now consider a limb-darkened source at $\vec y^{(0)}$ with radius $R_\mathrm{src}$,  
with the brightness profile
\begin{equation}
I_\mathrm{s}(\vec y) = f_\mathrm{I}\left(\frac{\left(\vec y - \vec y^{(0)}\right)^2}{R_\mathrm{src}^2}\,;
\,\tilde{u}
\right)\,,
\end{equation}
where
\begin{equation}
f_\mathrm{I}(r; \tilde{u}) = \frac{1}{1-\tilde{u}/3}\,
\left(\tilde{u}\,\sqrt{1-r^2}+1-\tilde{u}\right)
\end{equation}
for $ 0 \leq r \leq 1$, and $f_\mathrm{I}(r,\tilde{u}) = 0$ for $r > 1$.
This profile is normalized so that Eq.~(\ref{normalization}) is satisfied.
The parameter $\tilde{u}$ is chosen from the interval $[0,1]$, where $\tilde{u} = 0$ corresponds
to a uniformly bright source.
This brightness profile is the simple 'linear' model\footnote{The profile involves a linear term of $\cos\theta$, where
$\theta$ is the angle between the normal to the surface and the direction to the observer.} widely used, though for some type of stars, one should add
other terms (e.g. Claret et al.~\cite{Claret}). The general ideas
of the approach discussed here
do not depend on this special choice, especially terms like $(1-r^2)^{n}$ (i.e. like
$\cos^{2n} \theta$) can be treated in complete
analogy to the $\sqrt{1-r^2}$-term.\footnote{In fact, for the discussion of the MACHO 97-BLG-28 event 
(Albrow et al.~1998a,b\nocite{M28Planet1}\nocite{M28Planet2}), a $\sqrt[4]{1-r^2}$-term has also been included in the brightness profile.}

From Eqs.~(\ref{magn0}) and~(\ref{normalization}), one obtains the magnification as
\begin{equation}
\mu = \frac{1}{\pi R_\mathrm{src}^2}\,\int_{A_\mathrm{l}} I_\mathrm{l}(\vec x)\,
\mathrm{d}x_1\,\mathrm{d}x_2\,,
\end{equation}

To apply Green's theorem one has to find functions $P$ and $Q$ which satisfy
\begin{equation}
\frac{\partial Q}{\partial x_1} - \frac{\partial P}{\partial x_2} = I_\mathrm{l}(x_1,x_2)
= I_\mathrm{s}(\vec y(x_1,x_2))\,.
\end{equation}
Such functions are given by
\begin{eqnarray}
P(x_1,x_2) & = & -\frac{1}{2} \int_{x_2^{(0)}}^{x_2} I_\mathrm{s}\left(\vec y(x_1,x_2')\right)\,
\mathrm{d}x_2'\,, \\
Q(x_1,x_2) & = & \frac{1}{2} \int_{x_1^{(0)}}^{x_1} I_\mathrm{s}\left(\vec y(x_1',x_2)\right)\,
\mathrm{d}x_1'\,.
\end{eqnarray}
Since the brightness profile $I_\mathrm{s}$ and the lens equation $\vec y(\vec x)$ are given in
analytical form, the integrals can be evaluated numerically in general.

Since the brightness profile has an infinite slope at the limb of the source, it is advantageous to use
another continuation for $r > 1$.\footnote{As mentioned earlier, the value for $r>1$ does not play a role.}
Let us write the brightness profile in the form
\begin{equation}
f_\mathrm{I}(r; \tilde{u}) = C(\tilde{u}) \left(\tilde{u}\,B(r)+1-2\tilde{u}\right)\,,
\end{equation}
where
\begin{equation}
C(\tilde{u}) = \frac{1}{1-\tilde{u}/3}
\end{equation}
and
\begin{equation}
B(r) = \left\{\begin{array}{lcl} 1 + \sqrt{1-r^2} & \mbox{for} & 0 \leq r < 1 \\
				1 & \mbox{for} & r = 1 \\
			1-\sqrt{1-1/r^2} & \mbox{for} &  r > 1 \end{array}\right.\,.
\end{equation}
This definition is identical to the previous one except for $r > 1$.\footnote{One may also use
other continuations for $r>1$ which decrease differently for $r \to \infty$.}

The functions $P$ and $Q$ read
\begin{eqnarray}
P(x_1,x_2) &=& -C/2\,\Bigg[\left(1-2 \tilde{u}\right) (x_2-x_2^{(0)}) \,+ \nonumber \\
& & 
+\,\left. \tilde{u} 
\int_{x_2^{(0)}}^{x_2} B(r(x_1,x_2'))\,
\mathrm{d}x_2'\right] \,, \\ 
Q(x_1,x_2) &=& C/2\,\Bigg[\left(1-2 \tilde{u}\right) (x_1-x_1^{(0)}) \,+ \nonumber \\
& & +\,\left. \tilde{u} 
\int_{x_1^{(0)}}^{x_1} B(r(x_1',x_2))\,\mathrm{d}x_1'\right] \,,
\end{eqnarray}
where
\begin{equation}
r(x_1,x_2) = \frac{\left(\vec y(x_1,x_2) - \vec y^{(0)}\right)^2}{R_\mathrm{src}^2}\,.
\end{equation}
The function $B(r)$ has been chosen, so that for all $r$, $B(r) > 0$ in order to avoid contributions of different sign
in the numerical integration process.
The tail of the function is limited to 
\begin{equation}
\int_1^{\infty} B(r)\,dr = \frac{\pi}{2} - 1 \approx 0.571 \,,
\end{equation}
so that the integral is not dominated by the tail
contribution.

One remaining point of interest is the choice of the lower integration bound $\vec x^{(0)}$. 
To avoid unnecessary integration in the tail region of $B(r)$ that would result for
a fixed choice for $\vec x^{(0)}$  such as $(0,0)$, the lower integration bound can
be chosen as the center of each image which can be determined as shown in the next section.

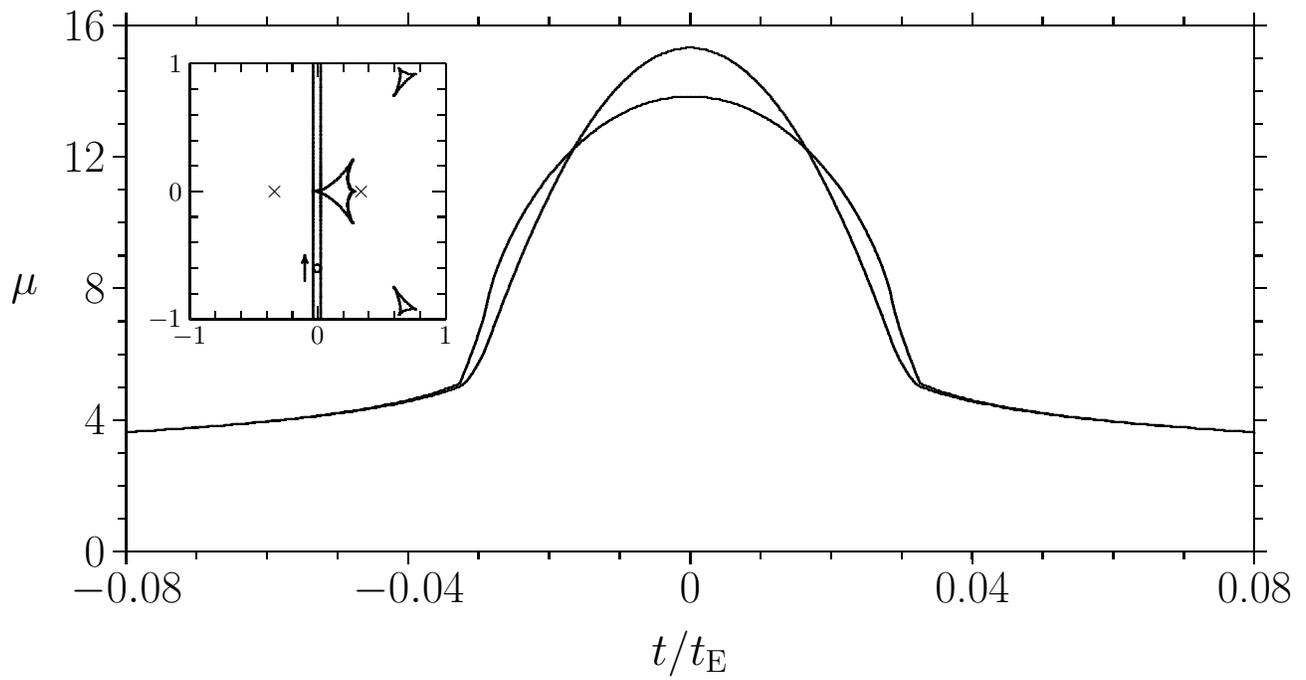
\begin{figure*}
\resizebox{12cm}{!}{\hspace{6cm}}
\hfill
\parbox[b]{55mm}{
\caption{The light magnification for a limb-darkened source with 
$\tilde{u} = 1$
 (curve with larger
peak amplification) and a uniformly bright source behind a binary lens. The angular source radius 
 corresponds to 0.03 $\theta_\mathrm{E}$, where $\theta_\mathrm{E}$ denotes the
angular Einstein radius\protect\footnotemark.
The mass ratio between the two lens components is 4, 
their angular separation is 0.68 $\theta_\mathrm{E}$. The source center passes
perpendicularly to the line connecting the lens objects at a minimal impact 
of 0.004 $\theta_\mathrm{E}$ from the midpoint towards
the lighter component.
$t_\mathrm{E}$ is the time in which the angular source position relative to the lens changes
by $\theta_\mathrm{E}$, $t = 0$ corresponds to
the point of time when the center of the source crossing the line connecting the lens components.
The inset shows the caustics, the position of the lens objects (crosses), and the size and
trajectory of the source which sweeps over a cusp caustic; the coordinates are multiples of 
$\theta_\mathrm{E}$.}
\label{lc}}
\end{figure*}

With these funtions $P$ and $Q$ one can approximate the magnification by the expression
\begin{eqnarray}
\mu = \frac{1}{2 \pi R_\mathrm{src}^2}\,\sum_{i=1}^{n} 
\left\{\left[Q\left(x_1^{(i)},\frac{1}{2}(x_2^{(i)}+x_2^{(i+1)})\right)\,+\right.\right. \nonumber \\
\left.\left.+\,Q\left(x_1^{(i+1)},\frac{1}{2}(x_2^{(i)}+x_2^{(i+1)})\right)\right]\,(x_2^{(i+1)}-x_2^{(i)})\right. + \nonumber \\
 +\,\left. 
\left[P\left(\frac{1}{2}(x_1^{(i)}+x_1^{(i+1)}),x_2^{(i)}\right)\,+\right.\right. \nonumber \\
 +\,\left.\left.P\left(\frac{1}{2}(x_1^{(i)}+x_1^{(i+1)}),x_2^{(i+1)}\right)\right]\,(x_1^{(i+1)}-x_1^{(i)}) 
\right\}\,,
\label{magnld}
\end{eqnarray}
which uses the symmetries and reduces to Eq.~(\ref{magnub}) with $\vec x^{(0)} = (0,0)$ 
for the case $\tilde{u} = 0$ (uniformly
bright source).
An example for a light curve for a limb-darkened source and one with a uniformly bright source behind a
binary lens is shown in Fig.~\ref{lc}.

\addtocounter{footnote}{-1}

\section{The centroid shift}
The position of the centroid of light can be determined in a similar way.
For $m$ images of the source, the position of the centroid of light $\vec x_\mathrm{c}$ is given by
\begin{equation}
\vec x_\mathrm{c} = \frac{\sum\limits_{j=1}^{m} \int_{A_\mathrm{j}} \vec x I_\mathrm{l}(\vec x)\,
\mathrm{d}x_1\,\mathrm{d}x_2}
{\sum\limits_{j=1}^{m} \int_{A_\mathrm{j}} I_\mathrm{l}(\vec x)\,
\mathrm{d}x_1\,\mathrm{d}x_2}
\label{centreq}
\end{equation}
where an integration has to be performed over the regions subtended by the images $A_\mathrm{j}$. The integral in the denominator
is just that discussed in the previous section and the integral in the numerator differs only by
an additional factor of $\vec x$. The integral can be solved in the way described in the last section
by only replacing $I_\mathrm{l}(\vec x)$ by $x_i\,I_\mathrm{l}(\vec x)$ ($i=1,2$).
For a uniformly bright source, the corresponding functions $P$ and $Q$ are
\begin{equation}
P(x_1,x_2) = -\frac{1}{2} x_1 x_2\,, \quad Q(x_1,x_2) = \frac{1}{4}x_1^2 
\end{equation}
for the $x_1$-component
and
\begin{equation}
P(x_1,x_2) = -\frac{1}{4} x_2^2\,, \quad Q(x_1,x_2) = \frac{1}{2} x_1 x_2
\end{equation}
for the $x_2$-component, so that one obtains with Eq.~(\ref{magnld}) the expressions
\begin{eqnarray}
x_{\mathrm{c},1} &=& \frac{1}{8 A}\,\sum_{i=1}^{n} 
\left\{({x_1^{(i)}}^2
+{x_1^{(i+1)}}^2)\,(x_2^{(i+1)}-x_2^{(i)})\right. + \nonumber \\
 & & +\,\left. 
({x_1^{(i)}}^2 -{x_1^{(i+1)}}^2)
(x_2^{(i+1)}+x_2^{(i)}) 
\right\}\,, \\
x_{\mathrm{c},2} &=& -\frac{1}{8 A}\,\sum_{i=1}^{n} 
\left\{({x_2^{(i)}}^2
-{x_2^{(i+1)}}^2)\,(x_1^{(i+1)}+x_1^{(i)})\right. + \nonumber \\
 & & +\,\left. 
({x_2^{(i)}}^2 +{x_2^{(i+1)}}^2)
(x_1^{(i+1)}-x_1^{(i)}) 
\right\}\,, 
\end{eqnarray}
where $A$ denotes the area of the image, which can be determined as shown in Sect.~\ref{secub}.

For a point source, the centroid's position can be written as
\begin{equation}
\vec x_\mathrm{c} = \frac{\sum_{j=1}^{m} \tilde{\mu}(\vec x_j)\,\vec x_j}{
\sum_{j=1}^{m} \tilde{\mu}(\vec x_j)}\,.
\end{equation}
If one compares this expression with that given by Eq.~(\ref{centreq}), one sees that the (finite)
source size has cancelled out in Eq.~(\ref{centreq}). 

\begin{acknowledgements}
I would like to thank P.~Sackett and K.~Sahu for reading and commenting on this manuscript.
\end{acknowledgements}

\footnotetext{
For a lens at a distance $D_\mathrm{d}$ and a source at a distance $D_\mathrm{s}$ from the observer, 
the angular Einstein radius for
a lens of total mass $M$ is given by $\theta_\mathrm{E} = \left(4GMc^{-2} D_\mathrm{d}^{-1} D_\mathrm{s}^{-1}(D_\mathrm{s} - D_\mathrm{d})
\right)^{1/2}$.} 

\clearpage
\LARGE

\vspace*{5cm}

Figure 1
\vspace*{1cm}

\epsfig{file=Ac162.f1}
\clearpage

\vspace*{5cm}

Figure 2
\vspace*{1cm}

\epsfig{file=Ac162.f2}

\end{document}